\documentclass[aps,prl,twocolumn,showpacs,groupedaddress,superscriptaddress]{revtex4}
\usepackage[latin1]{inputenc}
\usepackage{amsmath}
\usepackage{amsfonts}
\usepackage{amssymb}
\usepackage{graphicx}

\newcommand{\derp}[2]{\partial_{#1} #2}
\newcommand{\ga}{\Gamma}

\begin{document}
\title{Liquid films with high surface modulus moving in tubes: dynamic wetting film and jumpy motion}

\author{Isabelle Cantat}
\author{Benjamin Dollet}
\affiliation{Institut de Physique de Rennes, Universit\'{e} Rennes 1,
  UMR CNRS 6251, B\^atiment 11A, Campus Beaulieu,
   35042 Rennes Cedex, France
}

\begin{abstract}
We investigate the motion through a wet tube of transverse soap films, or lamellae, of high surface dilatationnal modulus. Combining local thickness and velocity measurements in the wetting film, we reveal a zone of several centimeters in length, the dynamic wetting film, which is significantly influenced by a moving lamella. The dependence of this influence length on lamella velocity and wetting film thickness provides a discrimination among several possible surfactant  minimal models. A spectacular jumpy mode of unsteady motion of a lamella is also evidenced.
\end{abstract}
\maketitle

The motion of a liquid meniscus sliding over a wet wall is ubiquitous in many industrial and biological contexts, among which dip coating \cite{degennes}, bubble or droplet motion in pores (e.g. in enhanced oil recovery) and microfluidic channels \cite{bruus}, foam friction on solid boundaries \cite{denkov05}, as well as lung diseases \cite{goerke98}.  The problem has first been studied by Landau and Levich  \cite{landau42} and Derjaguin \cite{derjaguin33} (LLD) for a solid plate pulled out of a liquid bath at small velocity. For pure liquids, the dynamics is controlled by the dynamic meniscus that forms between the static meniscus and the wetting film withdrawn by the plate.
The LLD model leads to a wetting film thickness $h_{\mathrm{LLD}} = 0.945 \ell_c \mathrm{Ca}^{2/3}$, with $\ell_c$ the capillary length and $\mathrm{Ca}= \eta U/\gamma$ the capillary number ($\gamma$: surface tension, $\eta$: liquid viscosity, and $U$: plate velocity); 
the extension of the dynamic meniscus along the plate scales like $L_{\mathrm{LLD}} \sim \ell_c \mathrm{Ca}^{1/3}$. These predictions are based on the lubrication approximation $h_{\mathrm{LLD}}/L_{\mathrm{LLD}} \ll 1$, and on a free shear boundary condition at the air/liquid interface.

If the liquid phase is a solution of surfactants, the interfacial stress depends on the surface concentration of surfactants $\Gamma$ and on the rheology of the surfactant layer. This stress reacts against the local area variation of the interface, with a dependence quantified by a surface dilatational modulus $E$. The friction against a wall of foams made of solutions with high $E$ is well predicted by the limiting case of a locally incompressible interface \cite{denkov05}: the whole interface then moves with the foam,
resulting into a strong shearing of the wetting film between the foam and the wall. For single bubbles and foams, the total interface area of each bubble is constant during the motion, which is compatible with the local incompressibility of the interface.

In other geometries, like dip coating or motion of a lamella (a soap film across a tube), the total area of each connected part of the interface varies, and the 
location and extension of the zone where area variations occur becomes  a central question. A classical assumption is that the area variations are localized in the static meniscus, and that, in contrast with the model by Denkov {\it et al.} \cite{denkov05}, the interface of the wetting film and of the dynamic meniscus move with the plate. The hydrodynamical problem is then governed by the same equations as for pure liquids, after a well adapted rescaling \cite{shen02}.
Several models go beyond these limiting cases, by including surfactant adsorption/desorption \cite{hirasaki85},  diffusion in the bulk \cite{ratulowski90} or an intrinsic surface viscosity \cite{scheid10}.
Recently, the full problem was numerically solved without the lubrication approximation, which allowed to include the whole meniscus into the simulation, with surface diffusion \cite{pozrikidis01} or surface diffusion and adsorption/desorption \cite{krechetnikov06}. However, no clear experimental evidence allows to discriminate between these different models.

Using a surfactant solution with high surface modulus and low bulk viscosity, we bring the first direct experimental evidence that a lamella moving in a wet tube can push the wetting film over centimetric distances, {\it i. e.} more than two orders of magnitude larger than  $L_{\mathrm{LLD}}$.  We thus introduce the concept of {\it dynamic wetting film}.
We show that its lateral extension, that we call the  {\it influence length}, depends on the initial film thickness and on the lamella velocity. Moreover, we show that the various processes that may govern the influence length lead to very different scalings, discriminated by our experimental data. 
Finally, we demonstrate a surprising jumpy behavior at high velocity, where the meniscus intermittently slips over its wetting film.

\begin{figure}[h]
\centerline{
\includegraphics[width=0.25\textwidth]{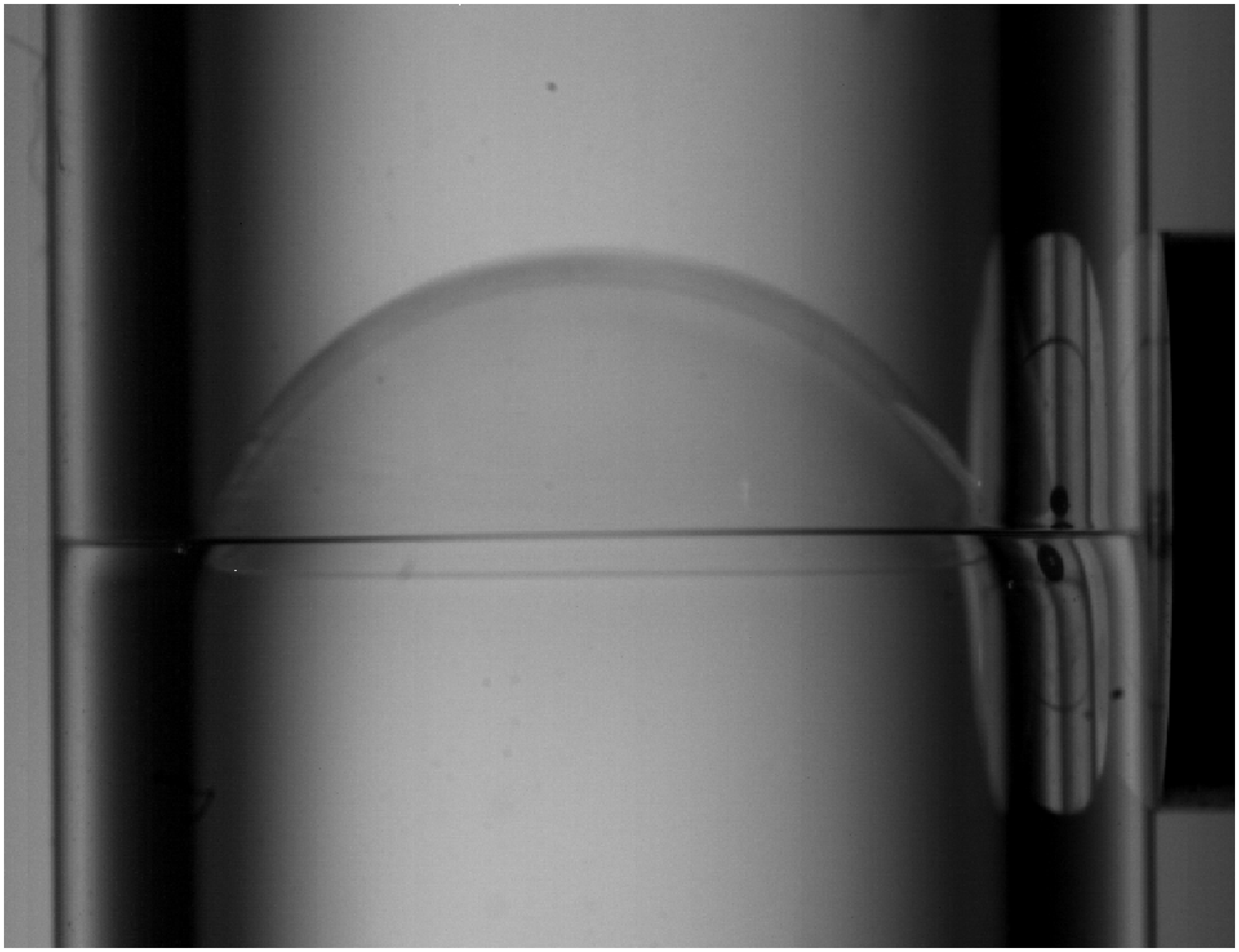}
\hspace*{0.2cm}
\includegraphics[width=0.14\textwidth]{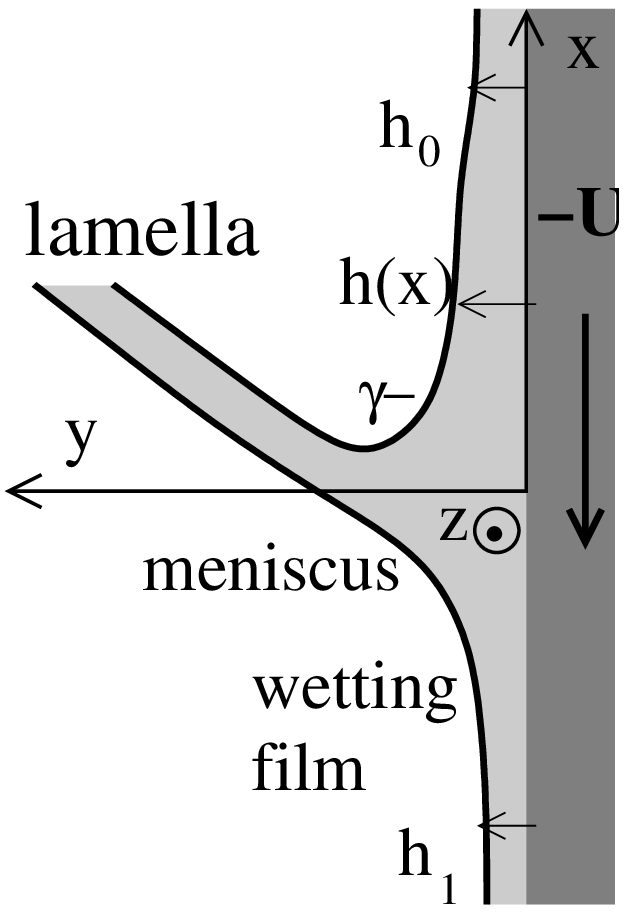}
}
\caption{(left) Lamella passing in front of the fiber. (right) Scheme of the meniscus and notations used in the text. }
\label{Fig:setup}
\end{figure}

Single lamellae are created and pushed at prescribed velocity $U$ in a wetted vertical tube of inner diameter $2a = 8.8$ mm following the method of \cite{dollet10}.
In order to maximize $E$, we use the well characterized mixture of sodium lauryl dioxyethylene sulfate, cocoamidopropyl betaine, and myristic acid (MAc) in ultrapure water (solution $S_1$), or the same with 40$\%$ wt glycerol (solution $S_2$), following the protocol of \cite{golemanov08}. The wetting film thickness $h$ is measured by white light interferometry using the commercial spectrometer USB4000 (Ocean Optics).
Neglecting multiple reflections within the film, the collected light intensity obeys
$I(\lambda) = I_0[ \alpha + \beta \cos(4 \pi n h/ \lambda)]$ with $I_0$ a reference intensity, $n=1.33$ the optical index of the solution and $\lambda$ the wavelength. The film thickness $h_{\mathrm{fib}}(t)$ is determined by computing the dominant Fourier component of $I(1/\lambda)$. 
Assuming a steady film profile in the lamella frame, we set $h(x) = h_{\mathrm{fib}}(t_0-x/U)$, with $t_0$ the time at which the lamella is in front of the fiber. The direction $x$ is oriented downstream, with $x=0$ at the lamella position (see Fig.\ref{Fig:setup}, right). 
A camera is synchronized with the spectrometer and records the film shape and velocity (Fig. \ref{Fig:setup}, left).  

\begin{figure}
\centerline{
\includegraphics[width=0.4\textwidth]{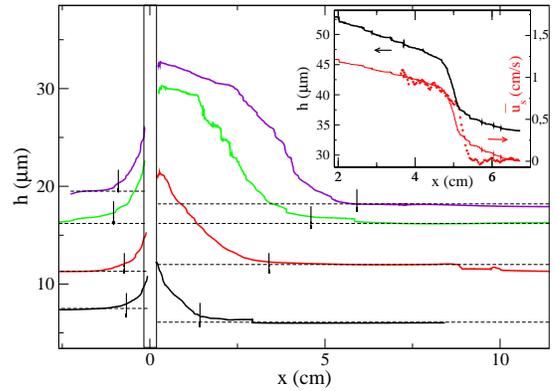}
}
\caption{Wetting film profiles obtained with different initial thicknesses and $U=0.6$ cm/s (solution $S_1$). The values obtained for $L_i$ and $h_i$ ($i=0,1$) after profile analysis are indicated resp. by thick segments and dashed lines. The central box corresponds to the meniscus and $x=0$ is the lamella position. Inset: Film profile obtained with $U= 1.7$ cm/s, $S_1$ (thick line), $\bar{u}_s$ measured by tracer tracking for the same experiment, using $\alpha= 0.45$ (dotted line) and $\bar{u}_s$ deduced from the film profile using Eq. \ref{us_de_h} (thin line).}
\label{Fig:profile}
\end{figure}

Typical wetting film profiles are given on Fig. \ref{Fig:profile}.
 The signal modulation is destroyed for too high thickness gradients, hence the meniscus profile is not accessible. The wetting film thickness $h_0$, far ahead the lamella, depends on the  waiting time since the previous lamella. Its evolution by drainage is slow enough for $h_0$ to remains uniform and constant on the experimental space and time scales, in the range 1 to 20 $\mu$m.  
 The thickness $h_1$ of the film deposited by the lamella does not need to be equal to $h_0$. If $h_0 \neq h_1$, the meniscus volume changes,
but slowly enough for the steady state assumption to remain valid.

\begin{figure}
\centerline{
\includegraphics[width=0.45\textwidth]{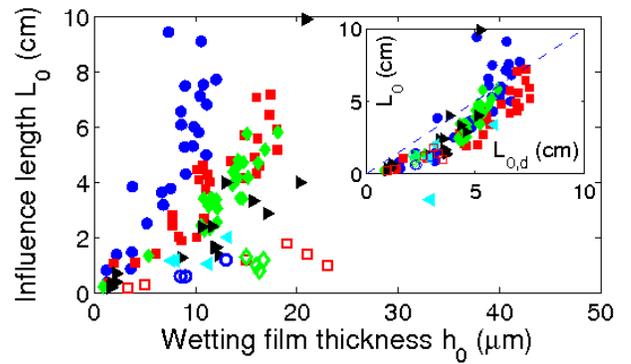} }
\caption{Influence length $L_0$ as a function of the wetting film thickness $h_0$ for two bulk viscosities (full symbols: $\eta_{S_1}=1.0$ mPa$\cdot$s, open symbols:  $\eta_{S_2},=4.0$ mPa$\cdot$s) and various velocities (in cm/s, $\circ, \bullet$: 0.32, $\square,\blacksquare$: 0.60, $\lozenge, \blacklozenge$: 1.15, $\triangleright, \blacktriangleright$: 2.35, $\triangleleft,\blacktriangleleft$: 4.20). Inset: $L_0$ as a function of  $L_{0,d}$ given by Eq. (\ref{linf_diffusive}). Used parameter values: $E_{S_1} =E_{S_2} = 0.1$~N/m \cite{golemanov08}, $D=10^{-10}$~m$^2$/s \cite{chang95}, $\Gamma_0 = 7\times 10^{-6}$~mol/m$^2$ (from the area per molecule $\approx$ 25~{\AA}$^2$ \cite{akamatsu91-petrov99}) and  $c_0 = c_{\mathrm{MAc}}= 0.88$~mol/m$^3$ \cite{golemanov08}, hence $h_\Gamma \approx 8$~$\mu$m.}
\label{Fig:Longueur_influence}
\end{figure}

Most remarkably, the wetting film begins to swell several centimeters in front of the lamella, at a distance $L_0$.
Some profile shapes are roughly exponential, but the thickest ones display a sharp transition between the flat wetting film and the swollen part of the film close to the meniscus, of typical thickness $h\approx 2 h_0$.
The rear influence length $L_1$ is significantly smaller, but still centimetric. 
We compute $L_0$ and $L_1$ using a systematic fitting procedure leading to the characteristic distances marked on Fig. \ref{Fig:profile}. $L_0$ is plotted as a function of $U$ and $h_0$ on Fig. \ref{Fig:Longueur_influence}.
The data are somewhat scattered, but exhibit a clear tendency: 
$L_0$ increases with $h_0$, and decreases with both $\eta$ and $U$; the best power law fit yields:
\begin{equation} \label{Eq:experimental_scaling}
L_0 \propto h_0^{1.0\pm 0.1} U^{-0.5\pm 0.1} \eta^{-1.2\pm 0.1} .
\end{equation}
A similar law was obtained for $L_1$ (data not shown).

In some experiments, the presence of tiny bubbles within the wetting film enabled the simultaneous measurement of the film thickness and of the surface velocity. These bubbles are convected at a velocity $\bar{u}_{b}= \alpha \bar{u}_s$, with $\bar{u}_s$ the surface velocity (the bar indicating velocities in the laboratory frame) and $\alpha \lesssim 1$ an unknown friction parameter. Comparison with the film thickness profile proves that, as the film begins to swell, the velocity rapidly increases from almost zero to a value of the order of $U$ (Fig. \ref{Fig:profile}, inset).
The velocity gradient in the dynamic wetting film is thus close to $U/h$, and the friction force per unit length (in the $z$ direction) can be estimated  as $f_v \approx \eta U (L_0/h_0 + L_1/h_1)$. 
This estimate, in the range 10 to 30 mN/m for all our experiments,  is in good qualitative agreement with the value extracted from the film images (Fig. \ref{Fig:setup}) $f_v = 2 \gamma_0 a/R$ \cite{dollet10}, with $\gamma_0 = 23.8$ mN/m the equilibrium surface tension \cite{golemanov08} and $R$ the radius of curvature of the lamella (data not shown). 
The friction in the dynamic meniscus alone is given by the Bretherton law $f_{v,B} = 4.70 \gamma_0 \mathrm{Ca}^{2/3}$ \cite{bretherton61} and is at most 1.6 mN/m in our range of velocities. It is thus negligible compared to the dynamic wetting film contribution for this solution. 

We also tested a SDS solution, with $\eta= 1.2$ mPa$\cdot$s and $E <$ 1 mN/m and a solution of 12-hydroxy stearic acid with ethanolamine as a counterion \cite{fameau11}, with $E= 38$ mN/m and a rheothinning behavior with a high viscosity $\eta$ varying between 20 and $10^{-2}$ Pa.s for shear rates between $10^{-2}$ and 300 s$^{-1}$. In both cases, no measurable influence length has been observed. This gives a strong hint that a large $L_0$  is associated to solutions of high $E$ and low $\eta$. The influence length results from a competition between the resistance of the surface against compression, and the viscous resistance of the bulk against shear.
The former may arise from various microscopic mechanisms, governed by the surface viscosity or elasticity, coupled with the surfactant desorption rate or diffusion. We now investigate them in their simplest form, at the expense of a quantitative modeling; however, this scaling approach appears to be sufficient to identify only one scenario able to capture the main dependences experimentally observed, in the investigated parameter range.

We  model the region $x>0$, ahead of the lamella, in the frame of the lamella. We consider a steady regime. Since $h\ll a$, the tube curvature is negligible and we assume an invariance in the $z$ direction. In the dynamic wetting film but at a distance larger than $L_{\mathrm{LLD}}$ from the lamella, Marangoni effects dominate over capillary effects and we can neglect the Laplace pressure. The velocity $u(x,y)$ is thus oriented along $x$ and varies linearly  between $-U$ at $y=0$ and $u_s(=\bar{u}_s-U)$ at $y=h$.
Mass conservation thus writes $0 = \partial_x [ (u_s -U) h/2]$, hence:
\begin{equation}
u_s= U \left(1 - \frac{2 h_0}{h}\right) , 
\label{us_de_h}
\end{equation}
in agreeement with experimental data (see Fig. \ref{Fig:profile},inset).

In the presence of the surface viscosity $\mu^*$, the tangential stress continuity imposes, at $y = h(x)$ \cite{edwards}, 
\begin{equation}
\eta \derp{y}{u}  =  \derp{x}{\gamma} +\mu^* \derp{xx}{u_s}\; .
\label{tang_stress}
\end{equation}
Ref. \cite{golemanov08} shows that without MAc, the surface modulus is two orders of magnitude lower. Therefore, it is reasonable to assume that $\partial_x \gamma$ is only related  to $\partial_x \Gamma_{\mathrm{MAc}}$, hence to use a one-component model.
Assuming for simplicity a linear relation for the surface tension $\gamma = \gamma_0 -E (\Gamma - \Gamma_0)/\Gamma_0$ with $\Gamma_0$ the equilibrium surface concentration in MAc and $\Gamma=\Gamma_{MAc}$, we get the velocity:
\begin{equation}
u(x,y) = \frac{1}{\eta} \left (\mu^* \derp{xx}{u_s} -\frac{E}{\Gamma_0}\, \derp{x}{\ga} \right) y   -U.
\label{v_de_h}
\end{equation}

If the interfacial dynamics was dictated by surface viscosity rather than surface elasticity, Eq. (\ref{v_de_h}) would become, for $y=h$ and $E=0$:
$1 + u_s/U =  (\mu^*h/\eta U) \partial_{xx} u_s$.
From experiments, the variation of $u_s$ is of the order of $U$ (Fig. \ref{Fig:profile}, inset), leading to the scaling:
\begin{equation}
L_{0,v} = \left (\frac{\mu^* h_0}{\eta}\right )^{1/2} ,
\label{linf_visc}
\end{equation}
which does not capture the observed dependence of $L_0$ on the lamella velocity.

We now neglect the interfacial viscosity to focus on the other contributions from the surfactants.
The surfactant mass balance at the interface writes:
\begin{equation}
\label{eq_set_dim1}
 \derp{x}{(u_s \ga) } = j  ,
\end{equation}
with $j$ the surfactant flux from the bulk to the interface. 
The dynamic wetting film is compressed by the moving lamella, inducing a typical surface concentration variation $\Delta\Gamma>0$. 
Moreover, Eq. (\ref{v_de_h}) leads to: $ u_s \sim -U - \frac{Eh}{\eta \Gamma_0} \derp{x}{\ga}$, from which we deduce $u_s \partial_x \Gamma \sim U \frac{\Delta \Gamma}{L_0}$ and $\Gamma \partial_x u_s \sim \frac{- \Delta \Gamma}{L_0} \, \frac{Eh_0}{\eta L_0}$. Experimentally, we get $\frac{Eh_0}{\eta UL_0} \approx 3$ with $h_0 \approx 10^{-5}$ m, $U\approx 1$ cm/s and $L_0 \approx 3$ cm,
showing, as expected, that the convective term $\partial_x(u_s \Gamma)$ is dominated by the velocity variation, which tends to accumulate surfactant in front of the lamella. The concentration gradient only reduces this effect. If we neglect this last term, we get the following scaling law for  the convective term in (\ref{eq_set_dim1}), as already obtained in \cite{hirasaki85}:
\begin{equation}
\partial_x ( \Gamma u_s) \sim - \frac{Eh_0 \Delta \Gamma}{\eta L_0^2} < 0. \label{scaling_convective_flux}
\end{equation}
This term is balanced by the exchange term with the bulk $j$. It equals the diffusive flux in the subphase, and obeys a kinetic sorption law \cite{edwards} that we assume linear. Then, $j = -D\partial_y c = k(c -  \Gamma/h_\Gamma)$ at $y=h$, with $c$ the bulk concentration in MAc, $c_0$ its equilibrium value, $h_\Gamma = \Gamma_0/c_0$,  $D$ 
the diffusion coefficient, and $k$ the sorption velocity, both assumed to be the ones of the micelles in which MAc is solubilized \cite{golemanov08}.
If $k$ is small, the exchange is limited by the desorption process, and $c\simeq c_0$. Then $j\sim -k\Delta\Gamma/h_\Gamma$, hence from (\ref{eq_set_dim1}) and (\ref{scaling_convective_flux}):
\begin{equation}
\label{linf_adsorp}
L_{0,a} \sim \left (\frac{E h_0  h_\Gamma}{\eta k} \right)^{1/2} .
\end{equation}
This scaling is exactly the same as in the viscous case (\ref{linf_visc}), with an effective surface viscosity $\mu^*_{\mathrm{eff}} = E h_\Gamma / k $.

If the desorption is fast, the surface is in equilibrium with the subphase and, close to the meniscus, the concentration increase in the subphase scales as $\Delta c(h) = \Delta \Gamma /h_\Gamma$. Then, the flux directly depends on the bulk diffusive field:
 $j = - D \partial_y c$. If the gradient is established over the diffusive distance $h_d\sim \sqrt{D L_0/U}$, this leads to the scaling behavior:
\begin{equation}
\label{linf_diffusive}
L_{0,d} \sim \left (\frac{Eh_{\Gamma}}{\eta} \right)^{2/3}  \frac{h_0^{2/3}}{(DU)^{1/3}} .
\end{equation}

This is only valid if $h_d < h_0$. Injecting (\ref{linf_diffusive}) in the expression of $h_d$, we get the condition $K=[ E D h_\Gamma/(\eta U^2 h_0^2)]^{1/3} < 1$. With the experimental parameter values (see caption of Fig. \ref{Fig:Longueur_influence}), we get $K \approx 2$.
If $h_d \gg h_0$, the gradient $\partial_y c$ becomes small and $j$ is estimated from surfactant conservation: $j = \partial_x (\int_0^h u c \mathrm{d}y) \simeq \partial_x (c\int_0^h u \mathrm{d}y)$.
From mass conservation, the flow rate $\int_0^h u \mathrm{d}y$ is a constant, equal to $-Uh_0$. Hence, $j\sim -Uh_0 \Delta c/L_0$, and $L_0$ in this convective regime is thus governed by:
\begin{equation}
\label{linf_conv}
L_{0,c} \sim \frac{E h_\Gamma}{\eta U} ,
\end{equation}
which does not capture the observed dependence of $L_0$ on $h_0$.
Overall, the only regime compatible with our experimental results is the diffusive one. Even if the exponents in (\ref{Eq:experimental_scaling}) and (\ref{linf_diffusive}) slightly differ, the rescaling by (\ref{linf_diffusive}) shown in Fig. \ref{Fig:Longueur_influence} (inset) provide a reasonable agreement with the experimental data, without adjustable parameter.
Moreover, we do not explain why $L_1$ is significantly shorter than $L_0$ (Fig. \ref{Fig:profile}); this suggests that the interface resists more compression than extension.

\begin{figure}
\centerline{
\includegraphics[width=0.35\textwidth]{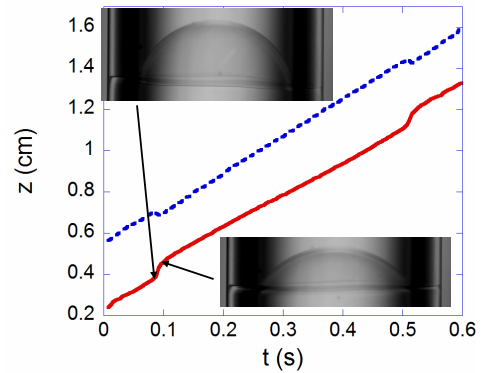} }
\caption{Time evolution of the position of the meniscus (plain line) and of the leading point of the lamella (dashed line), in the laboratory frame and in the jumpy regime. Two snapshots show the shape of the lamella just before, and just after, a jump.}
\label{Fig:Stick-slip}
\end{figure}

Finally, the most deformed lamellae deviate from steady motion by a striking scenario: they undergo periodic ``jumps" between phases of constant velocity (Fig. \ref{Fig:Stick-slip}). These jumps are quick (less than 10~ms) and macroscopic (of order 1~mm), which make them easily observable by naked eye. The curvature is partially released during jumps, and builds up again during the phases of constant velocity. Integrating Eq. (\ref{tang_stress}) (with $\mu^*=0$) along the front part of the dynamic wetting film yields $\gamma_- =\gamma_0 - \int_0^\infty \eta\partial_y u \mathrm{d}x \sim \gamma_0 - \eta U L_0/h_0$, with $\gamma_-$ the value of the surface tension close to the meniscus. The largest values obtained for $U L_0/h_0$ are close to $\gamma_0$, which shows that surface tension significantly decreases towards the meniscus. 
In this case, the linear relation between $\gamma$ and $\Gamma$ does not hold anymore, which may explain the discrepancy between the experimental and theoretical exponents.
It is likely that the interface becomes unstable below a certain value of $\gamma_-$ and collapses as in Langmuir monolayers \cite{lee08}, and that the jumps are macroscopic manifestations of such an instability. This unsteady behavior is in marked contrast with that of SDS films, which never showed such jumps \cite{dollet10}.

As a conclusion, our predictions of the length over which an air/liquid interface of a wetting film is entrained by a lamella may provide a quantitative criterion for the transition between tangentially immobile and mobile interfaces in foam/wall friction \cite{denkov05}. With $\ell$ a typical bubble size, these two limits correspond respectively to $L_0 \gg \ell$ and $L_0 \ll \ell$, and the transition criterion $L_0 = \ell$ may be expressed from our predictions, in terms of the material parameters of the surfactants. Experimental tests of this hypothesis are under way.

We thank A. Saint-Jalmes, N. D. Denkov and S. Tcholakova  for enlightening discussions about solution properties.


\end{document}